\documentclass[sigconf]{acmart}
\usepackage[utf8]{inputenc}
\usepackage{listings}
\usepackage{caption,subcaption}
\usepackage{booktabs,tabularx,makecell,multirow}
\usepackage{cleveref}

\lstdefinelanguage{graphql}{
	basicstyle=\ttfamily\scriptsize,
	string=[s]{"}{"},
	stringstyle=\color{red},
}

\graphicspath{{figures/}}

\setcopyright{iw3c2w3}
\copyrightyear{2020}
\acmYear{2020}
\acmISBN{}
\acmDOI{}

\acmConference[WWW '20]{WWW 2020}{April 20--24, 2020}{Taipei, Taiwan}

\begin{document}

\title{A Link Generator for Increasing the Utility of OpenAPI-to-GraphQL Translations}


\author{Dominik Adam Kus, István Koren, Ralf Klamma}
\orcid{0000-0003-1350-6732}
\affiliation{%
	\institution{RWTH Aachen University}
	\streetaddress{Informatik 5, Ahornstr. 55}
	\city{Aachen}
	\state{Germany}
	\postcode{52056}
}
\email{kus,koren,klamma@dbis.rwth-aachen.de}

\renewcommand{\shortauthors}{Developers Track}

\begin{abstract}
Standardized interfaces are the connecting link of today's distributed systems, facilitating access to data services in the cloud.
REST APIs have been prevalent over the last years, despite several issues like over- and underfetching of resources.
GraphQL enjoys rapid adoption, resolving these problems by using statically typed queries.
However, the redevelopment of services to the new paradigm is costly.
Therefore, several approaches for the successive migration from REST to GraphQL have been proposed, many leveraging OpenAPI service descriptions.
In this article, we present the findings of our empirical evaluation on the APIs.guru directory and identify several schema translation challenges.
These include less expressive schema types in GraphQL, as well as missing meta information about related resources in OpenAPI.
To this end, we developed the open source Link Generator, that analyzes OpenAPI documents and automatically adds links to increase translation utility.
This fundamentally benefits around 34\% of APIs in the APIs.guru directory.
Our findings and tool support contribute to the ongoing discussion about the migration of REST APIs to GraphQL, and provide developers with valuable insights into common pitfalls, to reduce friction during API transformation.
\end{abstract}

%



\maketitle

\section{Introduction}
The modern web is powered by APIs.
They enable all kinds of distributed applications, like web and mobile apps connecting to microservice architectures.
Prominent examples are cloud-oriented data processing frameworks running on the AWS Cloud or the Google Cloud Platform, and the orchestration of container applications with Kubernetes.
Since its proposal in the year 2000, the REST API style~\cite{Fiel00} has become the most used interface type on the web.
However, there are some shortcomings of REST-based interfaces.
In particular, \textit{over-} and \textit{underfetching} are common problems when designing such an interface~\cite{ZWLe19}.
Overfetching occurs when a client needs a specific piece of information but that information is only available from the API in conjunction with additional data.
The client has to fetch more data than it actually needs.
The opposite is true for underfetching: the client has to perform multiple requests against the interface to fetch all the information it requires.
This can lead to increased bandwidth usage and higher latency for the client, limiting the applications's performance and degrading the user experience.
Additionally, complex queries are particularly detrimental for data science in the context of high volume Internet of Things data, for instance in Industry 4.0 settings.
To tackle the shortcomings of REST, many new API-related technologies are currently being discussed.
One of those technologies is the modern web API style GraphQL.
After its release in 2015, it quickly gained popularity among developers and is used for many public APIs as of today.
Among others, GitHub and Facebook are using GraphQL to provide their services.
One reason for GraphQL's quick success is its promise to solve the above mentioned issues encountered when using REST APIs.
This is enabled by its query language which allows developers to write statically typed queries, thereby enabling clients to fetch complex data in a single request while also providing a high degree of granularity.
Together with being very developer-friendly, those properties make GraphQL highly interesting to providers of web services, in particular with mobile frontends.
As it can be very time consuming to migrate an existing REST interface to GraphQL or to maintain interfaces of both styles at the same time, there is a need for tools that can aid this transition.
One of those tools is the open-source project OpenAPI-to-GraphQL\footnote{\url{https://github.com/IBM/openapi-to-graphql}} introduced by Wittern et al.~\cite{WCLa18}.
It automatically generates GraphQL wrappers and a corresponding schema for existing REST APIs, based on their OpenAPI documentation.
Such a wrapper translates incoming GraphQL requests to the corresponding REST calls, issues those calls and assembles the responses into a valid GraphQL response.
Thereby, it acts as a middleman between the GraphQL client and the existing REST interface.
As a result, a wrapper works on top of an existing interface and does not required changes of the underlying application.

In this article, we discuss challenges that arise when creating a GraphQL wrapper, in particular using OpenAPI-to-GraphQL.
As the quality of the generated wrapper heavily depends on the source documentation, we statistically evaluate real-world API documentations in the public APIs.guru directory regarding their suitability for wrapper generation.
Finally, we introduce our link generator tool that aims to those documentations and therefore improve the wrapper generation process.

The article is organized as follows.
In the next section, we introduce related work in the area of API migration.
Section~\ref{sec:translation-challenges} then discusses schema translation challenges found in an empirical study.
Section~\ref{sec:link-generator} presents the implementation of the Link Generator.
Finally, Section~\ref{sec:conclusion} wraps up the article.

\section{Related Work}
Regarding the benefits GraphQL offers over REST, a case study conducted in~\cite{BMVa19} found a 99\% reduction in the size of a query result in the median when switching to GraphQL.
Aside from performance, the example in~\cite{Brya17} shows that GraphQL also has the potential to provide an outward data model that is similar to the internally used data model.
Furthermore, it enables fluid API development without the need for explicit versioning while still maintaining compatibility.

The concept of GraphQL wrappers as well as an implementation example on the client side is presented in~\cite{Lusc16}.
However, the wrapper is manually implemented in this example.
A tool for automatic wrapper generation is Swagger2GraphQL\footnote{\url{https://github.com/yarax/swagger-to-graphql}}.
It takes the Swagger 2.0 documentation of a REST interface as inputs and wraps the API automatically.
The tool OpenAPI-to-GraphQL which was introduces by Wittern et al.~\cite{WCLa18} works in a similar fashion but uses the more recent OpenAPI 3.0 format as input.
Therefore, it can leverage new features of the most recent OpenAPI version to enhance the generated GraphQL schema.
One of those is the link definition feature discussed in this article.

\section{Schema Translation Challenges}\label{sec:translation-challenges}
The first step to generate such a wrapper is translating the schema of the REST API into a GraphQL schema.
The schema determines the data model of the generated interface and therefore impacts the developer friendliness and usability of it.
However, the quality of the generated schema depends heavily on the source OpenAPI documentation.
To assess the influence of the source documentation on the generated wrapper, we performed an empirical evaluation of over 1500 of such documents found in the APIs.guru\footnote{\url{https://github.com/APIs-guru/openapi-directory}} directory.
In our analysis, we derived several translation challenges, particularly when using OpenAPI-to-GraphQL\footnote{OpenAPI-to-GraphQL version 2.0 was used during testing.}.

First, OpenAPI can model multiple different response types based on a HTTP \textit{success} status code (200–299).
This has no direct equivalent in GraphQL.
In such a case, OpenAPI-to-GraphQL displays a warning and translates only the numerically smallest defined status code, ignoring the others.
As a consequence, the generated wrapper may be missing some of the functionality of the original interface.
Around 26\% of APIs contain an at least one endpoint affected by this limitation.
Second, we compared the expression of data format descriptions in OpenAPI and GraphQL and found several specific properties in OpenAPI that cannot be expressed in a GraphQL schema, like \textit{patterns}.
OpenAPI schema definitions are based on the JSON Schema specification which allows narrowing down allowed values for data types~\cite{Open18}.
For example, they allow to define that a string has to be of a certain form using a \textit{pattern}.
The GraphQL schema language, however, does not contain an equivalent mechanism.
In total, we found that there are 16 such OpenAPI schema object properties with no direct GraphQL equivalent.
Thus, creating a wrapper for APIs using those properties will inevitably result in a loss of schema information.
An overview of those properties together with their prevalence in documentations found in the APIs.guru directory can be found in \Cref{fig:oai-schema-properties}.
While some of those properties, namely \textit{oneOf}, \textit{anyOf} and \textit{not}, are already being handled by OpenAPI-to-GraphQL at the time of writing, translating the others to GraphQL remains an open challenge.
In total, about 33\% of the documents include at least one of those properties.
\begin{table*}[t]
  \small
  \begin{tabularx}{\textwidth}{@{}lrrX@{}}
    \toprule
    \multirow{2}{*}{Property} & \multicolumn{2}{c}{Documents} & \multirow{2}{*}{Meaning}  \\
    \cmidrule{2-3}
                              & Count                         & Ratio                                &                                                                                                                                                                \\
    \midrule
    multipleOf                & 1                             & 0.1\%                                & Enforces that a number is a multiple of a given number.                                                                                   \\
    \addlinespace
    minimum                   & 287                           & 18.3\%                          & \multirow{2}{=}{Enforces that a number is within a given range.}                                                                                               \\
    maximum                   & 241                           & 15.3\%                           &                                                                                                                                                                \\
    \addlinespace
    exclusiveMinimum          & 1                             & 0.1\%                             & \multirow{2}{=}{Specifies whether the range is including or excluding the given minimum or maximum respectively.}                                              \\
    exclusiveMaximum          & 1                             & 0.1\%                               &                                                                                                                                                                \\
    \addlinespace
    minLength                 & 326                           & 20.8\%                          & \multirow{2}{=}{Enforces that a string has a length within a given range.}                                                                                     \\
    maxLength                 & 327                           & 20.8\%                          &                                                                                                                                                                \\
    \addlinespace
    pattern                   & 329                           & 20.9\%                          & Enforces that a string matches a given pattern.                                                                                                                 \\
    \addlinespace
    minItems                  & 110                           & 7.0\%                             & \multirow{2}{=}{Enforces that an array must have a number of items within a given range.}                                                                      \\
    maxItems                  & 111                           & 7.1\%                             &                                                                                                                                                                \\
    \addlinespace
    uniqueItems               & 12                            & 0.8\%                           & Enforces that an array contains no duplicate items.                                                                                                            \\
    \addlinespace
    minProperties             & 29                            & 1.8\%                             & \multirow{2}{=}{Enforces that an object must have a number of properties within a given range.}                                                                \\
    maxProperties             & 39                            & 2.5\%                              &                                                                                                                                                                \\
    \addlinespace
    oneOf                     & 7                             & 0.4\%                              & \multirow{2}{=}{Requires an object to adhere to exactly one or at least one schema within an array of schemas respectively.} \\
    anyOf                     & 2                             & 0.1\%                             &                                                                                                                                                                \\[\normalbaselineskip]
    \addlinespace
    not                       & 0                             & 0.0\%                               & Enforces that given data must not adhere to a given schema.                                                                                                    \\
    \bottomrule
  \end{tabularx}
  \caption{OpenAPI schema object properties with no direct equivalent in GraphQL and their prevalence on APIs.guru.}\label{fig:oai-schema-properties}
\end{table*}

Third, missing \textit{link definitions} in OpenAPI are a major pain point.
They allow defining inter-relationships between various resources, which are reflected in the translated GraphQL schema.
As a result, the added fields increase the readability and intuitiveness of the schema.
In some cases, especially when fetching a list of objects, it can even increase the amount of data that can be fetched in a single server roundtrip.
An example for the improvement achieved by adding link definitions is shown in \Cref{lst:github-query} using GitHub's v3 API.
It shows a query that fetches a repository and its list of branches from the wrapper generated by OpenAPI-to-GraphQL with and without using the link generator.
When links are added, a repository and its list of branches are no longer unrelated in the GraphQL API.
As a result, querying them is no longer done through two distinct top-level fields.
Instead, the branches become a property of their repository.
This is much more intuitive as conceptually, the branches are specific for the repository which is now reflected in the API schema.
Furthermore, it is no longer necessary for the client to specify the same parameters \texttt{owner} and \texttt{repo} twice in the query.
This is more concise and also eliminates a potential source of errors.
The example shows that adding links to an OpenAPI document can significantly improve the quality of the wrapper generated by OpenAPI-to-GraphQL.
However, despite their high importance for the translation to GraphQL, we could only find link definitions in 3 out of the over 1500 documents in the APIs.guru directory.
We see one reason for that in the current low adoption rate of OpenAPI 3.0.
As link definitions were first introduced with version 3 of the OpenAPI specification, documents using previous versions are inevitably missing link definitions.
To combat their low prevalence, we created \textit{OpenAPI-Link-Generator} to greatly improve generated GraphQL wrappers.
It automatically adds link definitions to OpenAPI documents whenever possible based on several heuristics.
In the process, it converts OpenAPI documents to version 3, if necessary.

\begin{figure*}
  \begin{minipage}{0.46\linewidth}%
    \subcaption{Without link generator}
    \begin{lstlisting}[language=graphql]
{
  repos(owner: "o", repo: "r") {
    name
  }
  reposBranches(owner: "o", repo: "r") {
    name
  }
}
    \end{lstlisting}
  \end{minipage}%
  \hfill
  \begin{minipage}{0.46\linewidth}%
    \subcaption{With link generator}
    \begin{lstlisting}[language=graphql]
{
  repos(owner: "o", repo: "r") {
    name
    branches {
      name
    }
  }
}
    \end{lstlisting}
  \end{minipage}%

	\Description{Querying a GraphQL wrapper for the GitHub API.}
  \caption{Querying a repository and its branches from the GraphQL wrapper generated by running OpenAPI-to-GraphQL on the GitHub API with and without using the link generator.}\label{lst:github-query}
\end{figure*}

\section{OpenAPI-Link-Generator}
\label{sec:link-generator}
\begin{figure*}
  \includegraphics[width=\linewidth]{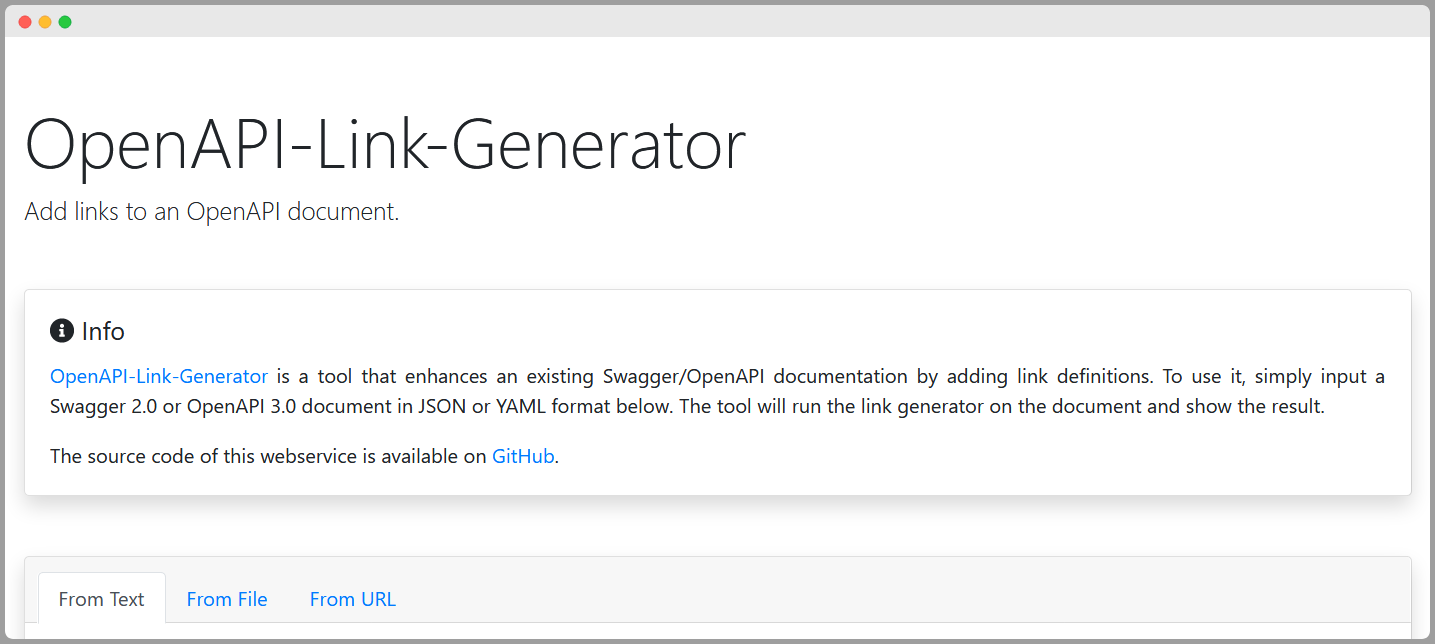}
  \Description{Screenshot of the OpenAPI-Link-Generator web service.}
  \caption{Screenshot of the OpenAPI-Link-Generator web service.}\label{fig:webservice-screenshot}
\end{figure*}

OpenAPI-Link-Generator was created to improve the quality of the transition to GraphQL.
To do that, it enables users to leverage the existing support for link definitions in OpenAPI-to-GraphQL without the need to manually go through the tedious process of adding links to existing API documentation.
As such, it can be used in a pre-processing step before applying OpenAPI-to-GraphQL to improve the translation result.
It is developed as open source software within our GitHub organization\footnote{\url{https://github.com/rwth-acis/openapi-link-generator}}.
The link generator is available as a freely accessible web service\footnote{\url{https://openapi-link-generator.herokuapp.com}} and as a Node.js library distributed through npm\footnote{\url{https://www.npmjs.com/package/openapi-link-generator}} which also includes a command line tool.
\Cref{fig:webservice-screenshot} shows a screenshot of the web application.
It provides three input possibilities.
First, the raw text can be pasted in.
Second, an OpenAPI file can be chosen from the local file system.
Third, a URL can be pasted it that links to a specification file.
After the link generation process, we show the diff and some statistics on the number of generated links.

It is noteworthy that we currently only consider HTTP GET requests for adding links.
For simplicity, we therefore use OpenAPI paths and the GET operation defined on that path synonymously in the following.
As there is no universal rule where a link should be added to an OpenAPI document and where not, the link generator has to rely on heuristics to add links.
The most important heuristic is the assumption that the forward slash in a URI implies a hierarchical relationship.
This is a general rule of REST API design according to~\citet{Mass12} and should therefore hold for most APIs.
Based on this assumption, the link generator tries to add links for each pair of paths of the form \texttt{/A} and \texttt{/A/B} where \texttt{A} and \texttt{B} may be arbitrary URI parts.
After such a path pair is identified, it is necessary to compare the parameters defined for those two paths.
Here, another heuristic is used.
If both operations contain a parameter with the same name as well as an identical schema, we assume that those two parameters have the same semantic meaning.
In this case, the link generator adds a parameter mapping for those two parameters to the generated link.
This is required to prevent the user from having to redefine the same parameters multiple times when fetching nested data.
Note that for this functionality, the link generator also parses internal references in the OpenAPI document according to the specification~\cite{Open18,BrZy12,BrZy12b}.

To evaluate the link generator, we ran it on the OpenAPI documentations found in the APIs.guru dataset.
In total, it added over 7500 link definitions, affecting about 34\% of the documents.
One prominent example for the improvement achieved by the link generator is shown in \Cref{lst:github-query} and is discussed in \Cref{sec:translation-challenges}.
We are convinced that our tool is a valuable addition to the growing ecosystem of web API tooling, enabling a more sustainable translation from OpenAPI to GraphQL.

\section{Conclusion}
\label{sec:conclusion}
GraphQL is one of the most promising new API styles, rapidly gaining popularity and market share.
It provides benefits for developers and providers alike and consequentially is appealing to many providers of REST APIs.
The open-source tool OpenAPI-to-GraphQL can be used to automatically generate a GraphQL API from an existing REST interface by creating a wrapper based on its OpenAPI documentation.
However, missing link definitions in OpenAPI documents limit the quality of the generated wrapper.
To tackle this problem, we proposed the OpenAPI-Link-Generator tool.
It can be used to automatically add link definitions to existing OpenAPI documentations and thereby improve the outcome of a GraphQL migration.
The tool is available through npm\footnote{\url{https://www.npmjs.com/package/openapi-link-generator}} as a nodejs module and as a command line tool.
Furthermore, we provide a web-based tool\footnote{\url{https://openapi-link-generator.herokuapp.com}} that can apply the link generator to any OpenAPI document at the click of a button.

\begin{acks}
Funded by the Deutsche Forschungsgemeinschaft (DFG, German Research Foundation) under Germany's Excellence Strategy – EXC-2023 Internet of Production – 390621612.
\end{acks}

\bibliographystyle{ACM-Reference-Format}
\bibliography{linkgenerator}

\end{document}